\input harvmac
\input epsf
\def\figin{\epsfcheck\figin}\def\figins{\epsfcheck\figins}
\def\epsfcheck{\ifx\epsfbox\UnDeFiNeD
\message{(NO epsf.tex, FIGURES WILL BE IGNORED)}
\gdef\figin##1{\vskip2in}\gdef\figins##1{\hskip.5in}
\else\message{(FIGURES WILL BE INCLUDED)}%
\gdef\figin##1{##1}\gdef\figins##1{##1}\fi}
\def\DefWarn#1{}
\def\figinsert{\goodbreak\midinsert}
\def\ifig#1#2#3{\DefWarn#1\xdef#1{fig.~\the\figno}
\writedef{#1\leftbracket fig.\noexpand~\the\figno}%
\figinsert\figin{\centerline{#3}}\medskip\centerline{\vbox{\baselineskip12pt
\advance\hsize by -1truein\noindent\footnotefont{\bf Fig.~\the\figno:} #2}}
\bigskip\endinsert\global\advance\figno by1}

\overfullrule=0pt
\def\Title#1#2{\rightline{#1}\ifx\answ\bigans\nopagenumbers\pageno0\vskip1in
\else\pageno1\vskip.8in\fi \centerline{\titlefont #2}\vskip .5in}

\lref\gh{G. W. Gibbons and S. W. Hawking, {\it Action integrals and
partition
functions in quantum gravity},  Phys. Rev. {\bf D15} (1977)
2752.}
\lref\cw{C. Callan and F. Wilczek, {\it On geometric entropy},
  Phys. Lett. {\bf B333} (1994) 
55; hep-th/9401072.}
\lref\ct{S. Carlip and C. Teitelboim, {\it The off shell black hole}
gr-qc/9312002.}
\lref\mark{M. Srednicki, Phys. Rev. Lett. {\bf 71} (1993) 666,
{\it Entropy and area}, hep-th/9303048.}
\lref\bkls{L. Bombelli, R. Koul, J. Lee 
and R. Sorkin, { \it A quantum source of entropy for black holes},
 Phys. Rev. {\bf D34} (1986) 373.} 
\lref\gub{S. S. Gubser, {\it AdS/CFT and gravity},   hep-th/9912001.}
\lref\hhu{S. Hawking and C. Hunter, {\it Gravitational entropy and global
structure}, Phys. Rev. {\bf D59} (1999)044025,  hep-th/9808085.}
\lref\rs{L. Randall and R. Sundrum, {\it An alternative to
compactification},
Phys. Rev. Lett. {\bf 83} (1999) 4690,  hep-th/9906064.  }
\lref\su{L. Susskind and J. Uglum, {\it Black hole entropy in 
canonical quantum gravity and superstring theory}, Phys. Rev. {\bf D50}
(1994) 2700,  hep-th/9401070.}
\lref\kb{D. Kabat, {\it Black hole entropy and entropy of entanglement}, 
Nucl. Phys. {\bf B453} (1995), hep-th/9503016.}
\lref\myers{J-G. Demers, R. LaFrance and R. Myers, {\it Black hole entropy 
and renormalization},  gr-qc/9507042.}
\lref\jm{J. Maldacena, {\it The large N limit of superconformal field
theories
and supergravity}, Adv. Theor. Math. Phys. {\bf 2} (1998) 231,
hep-th/9711200. }
\lref\hs{M. Henningson and K. Skenderis, {\bf JHEP} 9807 (1998)
023, hep-th/9806087.}
\lref\vf{P. Frampton and C. Vafa, {\it Conformal approach to particle
phenomenology}, hep-th/9903226.  }
\lref\bk{V. Balasubramanian and P. Kraus, {\it A stress tensor for anti de
Sitter gravity}, Commun. Math. Phys. (1999) 208, hep-th/9902121.}
\lref\bh{J. D. Brown and M. Henneaux, {\it Central charges in the canonical
realization
of asymptotic symmetries: An example from three dimensional gravity}, 
  Commun. Math. Phys. (1986), 104.}
\lref\fpst{T. Fiola, J. Preskill, A. Strominger and S. P. Trivedi,
{\it Black hole thermodynamics and information loss in two-dimensions},
Phys. Rev. {\bf D50} (1994) 3987,
hep-th/9403137. }
\lref\tj{T. Jacobson, {\it Black hole entropy and induced gravity}, 
 gr-qc/9404039.}
\lref\ch{C. Holzhey, F. Larsen and F. Wilczek, {\it Geometric and
renormalized
entropy in conformal field theory},  Nucl. Phys. {\bf B424} (1994) 443,
 hep-th/9403108.}
\lref\ascv{A. Strominger and C. Vafa, {\it Microscopic origin of the
Beckenstein-Hawking entropy}, Phys. Lett. {\bf B379} (1996) 99,
hep-th/9601029. }
\lref\hver{H. Verlinde, {\it Holography and compactification}, 
hep-th/9906182.  }
\lref\ffz{V. Frolov, D. Fursaev and A. I. Zelnikov, {\it Statistical 
origin of black hole entropy in induced gravity}, 
Nucl. Phys. {\bf B486} (1997) 339, hep-th/9607104.  }
\lref\ff{V. Frolov and D. Fursaev, {\it Mechanism of generation of black
hole
entropy in Sakharov's induced gravity}, 
Phys. Rev. {\bf D56} (1997) 2212, hep-th/9703178.}
\lref\lw{F. Larsen and F. Wilczek, {\it Renormalization of black hole
entropy and of the gravitational coupling constant}, 
 Nucl. Phys. {\bf B58} (1996) 249,
hep-th/9506066.}
\lref\dvv{J. DeBoer, E. Verlinde. and H. Verlinde, {\it On the  holographic
renormalization group},  hep-th/9912012.  }
\lref\gk{S. Gubser and I. Klebanov, {\it Absorption by branes and Schwinger
terms in the worldvolume theory}, Phys. Lett. {\bf B413} (1997) 41, 
 hep-th/9708005.}
\lref\suwi{L. Susskind and E. Witten, {\it The holographic bound in
 Anti-de-Sitter space}, hep-th/9805114.}
\lref\pp{A. Peet and J. Polchinski, {\it UV/IR relations in AdS dynamics},
Phys. Rev. {\bf D59} (1999) 065011, hep-th/9809022. }
\lref\witten{ E. Witten, 
{\it New dimensions in field theory and string theory}, 
{\rm  http://www.itp.ucsb.edu/online/susy${}_{}$c99/discussion}.}
\lref\tjj{T. Jacobson, {\it On the nature of black hole entropy}, 
gr-qc/9908031.}

\def\p{\partial}
\Title{\vbox{\baselineskip12pt
\hbox{hep-th/0002145}\hbox{}}}
{\vbox{\centerline {DeSitter Entropy, }
\centerline {Quantum Entanglement and AdS/CFT} }}
\centerline{Stephen Hawking$^*$, Juan Maldacena$^\dagger$ and
Andrew Strominger$^\dagger$ }
\bigskip
\centerline{ $^*$Department of Applied Mathematics and Theoretical Physics,}

\centerline{Centre for Mathematical Sciences}
\centerline{Wilberforce Road, Cambridge CB3 OWA, UK}
\bigskip\centerline{$^\dagger$Department of Physics}
\centerline{Harvard University}
\centerline{Cambridge, MA 02138,   USA}

%
%
\centerline{\bf Abstract}
{A deSitter brane-world bounding regions of anti-deSitter space
has a macroscopic entropy given by one-quarter the area of the
observer horizon.  A proposed variant of the AdS/CFT correspondence
gives a dual description of this cosmology as conformal field theory
coupled to gravity in deSitter space. In the case of two-dimensional
deSitter space  
this provides a microscopic
derivation of the entropy, including the one-quarter, as
quantum entanglement of the conformal field theory across the horizon.}

\smallskip
\noindent
\Date{}
\listtoc
\writetoc

\newsec{Introduction}
Despite advances in the understanding of black hole entropy,
a satisfactory microscopic derivation of the entropy of deSitter space \gh\
remains to be found. In this paper we address this issue in the context
of a deSitter space arising as a brane-world of the 
type discussed by  Randall and Sundrum \rs\ \foot{Unlike 
\rs\ we include a nonzero cosmological constant on the brane.} 
bounding two regions of
anti-deSitter space.
It is natural to suppose that such
theories are dual, in the spirit of AdS/CFT \jm , to a conformal
theory on the brane-world coupled to gravity with
a cutoff. The cutoff scales with the deSitter radius in such a way that the
usual AdS/CFT correspondence is recovered when the cutoff is taken to
infinity.  This duality provides an alternate description of the
deSitter cosmology which can be used, in the case of two 
dimensions, for a microscopic derivation of the
deSitter entropy. We find that the entropy can be ascribed to the quantum
entanglement of the CFT vacuum across the deSitter horizon. Quantum 
entanglement entropy can also be viewed as the entropy of the thermal
Rindler particles near the horizon, thereby avoiding reference to 
the unobservable region behind the horizon. 

Our derivation is closely related to the observation of reference 
\fpst\ that in two
dimensions black hole
entropy can be ascribed to quantum entanglement if
Newton's constant is wholly induced by quantum fluctuations of ordinary matter
fields (see also \refs{\tj, \su, \ffz, \ff}). 
In the context of  \fpst\ this seemed to be a
rather artificial and unmotivated assumption. However
the AdS brane-world scenarios do appear to have this feature.
The basic reason is that, in a semiclassical expansion, 
 the Einstein action on the brane
arises mainly from bulk degrees of
freedom\foot{ This follows when
the AdS radius is large compared to the Planck length.}
which correspond, in the dual picture, to ordinary matter fields on the
brane.
The semiclassical expansion in the bulk corresponds to a large $N$ 
expansion in the brane, in which the leading term in 
Newton's constant is induced by matter fields. 

Our derivation of two-dimensional deSitter entropy 
is similar to the derivation of black hole entropy
in \ascv\ in that it uses a brane field theory dual to the
spacetime gravity theory to compute the entropy. However it
differs in that in \ascv\ the black hole entropy was given by the
logarithm of the number of unobserved microstates of the black hole, 
whereas here the deSitter entropy arises from
entanglement with the unobserved states behind the
horizon.\foot{In general the entanglement
entropy is less than the logarithm of the number of possible
microstates of the unobserved sector of the Hilbert space. If
the total system is in a pure state, the entanglement entropy is
bounded from above by the logarithm of the number of possibly
entangled states in the unobserved sector of the Hilbert space.} Alternatively, it can be viewed as the number 
of microstates of the thermal 
gas of Rindler particles  near the horizon.  This latter viewpoint is 
closer to that of \ascv.  This issue is 
explored  in the final section by throwing a 
black hole in the bulk of AdS at the brane. When the bulk black
hole reaches the brane, the brane state collapses to a brane black hole. 
At all stages the entropy is accounted for 
by a thermal gas on the brane.

Formally, the derivation can be generalized to higher-dimensional 
deSitter spaces which bound higher-dimensional anti-deSitter spaces. 
It was conjectured by Susskind and Uglum 
\su\ that there is a general a precise relationship 
between entanglement entropy and the one loop correction to Newton's
constant. Based on this, Jacobson \tj\ argued that black hole
entropy can be ascribed to quantum entanglement if
Newton's constant is wholly induced. However, 
while we are sympathetic to the conjecture of \su, and it fits well 
with the discussion herein, its status remains unclear \refs{\kb, \tjj}. 
The basic problem is that in greater than two dimensions the 
corrections have power law divergences and hence are regulator 
dependent. This makes precise statements difficult above two dimensions.    

A further 
significant fly in the ointment - even in two dimensions- is that there is no
known example of the type of brane-world scenario considered in
\rs\
embedded in a fully consistent manner into string theory \foot{
See however \hver\ for related scenarios in string theory.}.
The observations of the present paper are relevant only
if such examples exist. For the time being however they provide
intriguing connections along the circle of ideas pursued in
\refs{\gh-\ascv}.

\newsec{Classical Geometry}

\ifig\feuclidean{
Euclidean instanton geometry. The brane is an $S^2$ which bounds two
patches of euclidean $AdS_3 = H^3$.
}{\epsfxsize1.0in\epsfbox{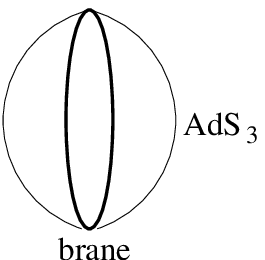}}

The euclidean action for a onebrane coupled to gravity with a negative
cosmological constant ($\Lambda=-{1 \over L^2}$) is
 \eqn\sac{S_{tot}=-{M_p \over 16 \pi}\int d^3x\sqrt{g}(R+
{2 \over L^2})+{T }\int d^2 \sigma
 \sqrt{h}.}
 $M_p$ here is the three dimensional Planck mass,
$T$ is the brane tension and $h$ is the induced metric on the
 brane. We have assumed that there is no boundary.
 We wish to consider a spherically symmetric brane at radius $r_B$
 which bounds two regions of $AdS_3$ with metrics
 \eqn\asgm{ds^2_3 =L^2dr^2 +L^2\sinh^2rd\Omega_2^2,}
 where $0\le r \le r_B$, as shown in \feuclidean .
 The two copies of $AdS_3$ are glued together along
 the $S^2$ at
$r=r_B$ where the bulk curvature has a delta function. The topology of
the spacetime is $S^3$.
 The induced metric on the brane is
 \eqn\bgm{ds^2_2 =\ell^2 d\Omega_2^2,}
with
\eqn\ldf{\ell \equiv L \sinh r_B.}
 The action for such a configuration is
 \eqn\fry{S_{tot}={M_p \over4 \pi L^2}V_3-{M_p\coth r_B \over 2 \pi L}
V_2 +{T } V_2,}
 where $V_3$ is the bulk volume and $V_2$ is the brane volume.
The second term arises from a delta function in the bulk curvature at
$r= r_B$.
One finds using \asgm\ that
 \eqn\tryu{S_{tot}=-{LM_p \over 2 }(\sinh 2r_B+2r_B)+ 4 \pi TL^2
 \sinh^2 r_B .}

The action \tryu\ has an extremum at
\eqn\trl{ \tanh r_B = {M_p \over 4 \pi TL}}
for which
\eqn\sfa{S_{tot}=-LM_p r_B.}
We are interested in the case that the right hand side of \trl\ is close to
(but less than) one so that $r_B$ and $\ell$ are large.
We may then approximate
\eqn\sfas{S_{tot}=-LM_p \ln {\ell }+ \dots }
The subleading corrections are suppressed for large $\ell$.

The induced brane metric
\bgm\ is the two-dimensional euclidean deSitter ($i.e.$ round $S^2$) metric
with a large
deSitter radius $\ell $.
A lorentzian deSitter solution can be obtained by
analytic continuation of the periodic angle $\phi$ on $S^2$ to
$it$. One finds
\eqn\rtzy{\eqalign{ds^2_3&=L^2dr^2+L^2\sinh^2r(d\theta^2-sin^2\theta
dt^2),\cr
ds^2_2&=\ell^2{d\theta^2-\ell^2 sin^2\theta dt^2}.}}
\ifig\fpenrose{ 
Penrose diagram of Lorentzian two dimensional de-Sitter space. The dotted
line
indicates the trajectory of a geodesic observer.  We have also indicated
the past and future horizons for that observer,
and the shaded region indicates
the patch covered by the coordinates \rtzy .
}{\epsfxsize2.0in\epsfbox{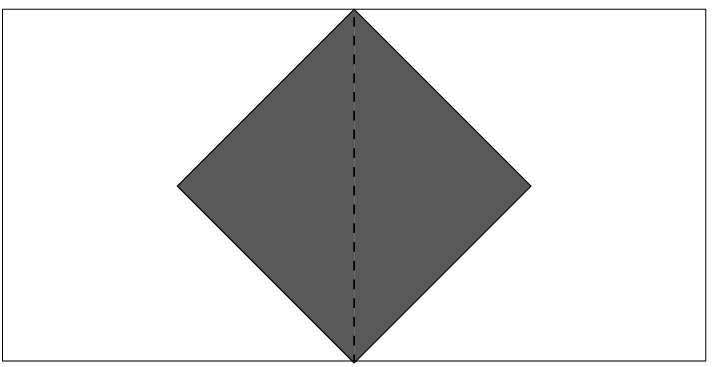}}
These coordinates cover the diamond-shaped region of $DS_2$
illustrated in \fpenrose . It is the region outside both
the future and past horizons of any timelike
observer at constant $\theta \neq 0, \pi$.

\newsec{Dual Representations}

Let us assume there is a unitary  quantum theory
whose semiclassical gravitational dynamics is  described by \sac.
Such a theory should have two dual
descriptions.\foot{Related discussions
appear in \refs{\gub,\dvv}.} The first is, as
described above, a three-dimensional bulk theory containing
gravity and a brane.

The second description is as a two-dimensional effective
theory of the light fields on the brane worldvolume.
These light fields include holographic matter living on the
brane. To see this we first consider a single copy of
$AdS_3$ in coordinates \asgm\ with a fixed boundary at $r=r_B$,
for large $r_B$.
If we keep the metric on the boundary fixed, but integrate over
the bulk metric, the resulting theory has a holographic description as
a $1+1$ conformal field theory on a sphere of radius
$\ell$ with central charge $c={3 LM_p \over 2}$ \bh\
and a cutoff at $L$\refs{\jm,\suwi,\pp}.\foot{Alternately one may take a sphere of
unit radius  and a cutoff at $L/\ell$.}
To recover the
geometry under consideration, we must take two copies of such bounded
$AdS_3$ spacetimes, identify them along their boundaries, and then integrate
over boundary metrics. Because there are two copies, one has two
copies of the matter action on the boundary, with a total central charge
\eqn\ccrg{c=3LM_p.}
The second brane tension term in the action \sac\ corresponds to
a counterterm which renormalizes the cosmological constant.

We are not fixing the boundary by hand so 
we have
two-dimensional
gravity because, as shown in \rs, there is a graviton zero mode
trapped on the brane, as well as the radion field representing the
radial position
of the brane. The radion has a small mass in the case of a large deSitter
boundary. For a $D$-dimensional boundary brane, gravity plus the radion
comprise $\half(D^2-3D+2)$ local degrees of freedom (after implementing the
constraints). Hence for the present case of $D=2$ the radion-gravity system
has no local degrees of freedom.

For our purposes we need only the gravity
part of the effective action, with the radion field set at
the minimum of its potential. The gravity effective action
is most easily represented in conformal gauge
\eqn\drl{ds_2^2=e^{2\rho}d\hat s_2^2,}
where $d\hat s^2$ is the unit metric on $S^2$ obeying
\eqn\rhp{\hat R_{z \bar z}=\hat g_{z \bar z},~~~~~~~~~~\int d^2 z
\hat g_{z \bar z}=4\pi }
in complex coordinates.
One then finds\foot{This is equivalent to the computation in
\refs{\hs,\bk}.}
\eqn\sacy{S_{g}=-{LM_p \over 4 \pi}
\int d^2 z\bigl(\p_z \rho \p_{\bar z}\rho +
\hat R_{z \bar z}\rho -{1 \over 2\ell^2} \hat g_{z \bar
z}
e^{2\rho} \bigr).}
The equations of motion for constant fields give
\eqn\eft{~~~~~~~~~~~\rho=\ln \ell.}
The action \sacy\ evaluated at this solution agrees with \sfas\
We note also that the total gravity plus matter central
charge vanishes, as required for general covariance.
These considerations determine \sacy.

\newsec{DeSitter Entropy}

In this section we give macroscopic and microscopic derivations of the
entropy.
\subsec{Semiclassical Macroscopic Entropy}

The macroscopic entropy can be computed directly in three dimensions
from the area entropy-law
\eqn\sfjt{S_{dS}={Area \over 4G_3}.}
In this expression $G_3=1/M_p$ and the horizon area is the area of the
fixed point of a $U(1)$ isometry \hhu\ of the instanton geometry
\asgm. This consists of a geodesic circle intersecting the
$S^2$ brane at the north
and south poles. The area (length) of this circle is
$4Lr_B \sim 4L \ln \ell$. Hence we obtain
\eqn\wki{S_{dS}= LM_p\ln \ell .}

An alternate derivation can be given from the
two dimensional deSitter space using
\eqn\sfj{S_{dS}={Area \over 4G_2}.}
The area in this formula is just the area of the observer horizon
($\theta=0,\pi$ in \rtzy\ ) which consists of two points and is therefore
equal to 2.  $G_2$ is determined as the (field-dependent) coefficient of
the the scalar curvature $R=\half g^{z \bar z}R_{z \bar z}$.
{}From \sacy\ this is
\eqn\fyj{{ 1 \over G_2}=2LM_p \rho =2 LM_p \ln \ell.}
Inserting \fyj\ into \sfj\ reproduces
\wki.

\subsec{Microscopic Entropy}
Let us now consider the entropy from the point of view of the
brane CFT with $c=3M_pL$. An $SO(2,1)$ invariant vacuum for quantum field theory in
lorentzian deSitter space $|0 \rangle $ can be defined as the
state annihilated by positive frequency modes in the metric
\eqn\dss{ds_2^2=\ell^2{-dt^2+dx^2 \over cos^2 t } .}
The proper time $\tau$
of an observer moving along a geodesic at
$x=0$ is related to the time $t$ in \dss\ by  time $e^{\tau/\ell} =
\tan({t\over 2}+{\pi \over 4})$.  Green functions in this vacuum
are single valued functions of $t$. Therefore they are periodic in
imaginary $\tau$ with period $2\pi i \ell$, and the observer accordingly
detects a thermal bath of particles with temperature $1 \over 2 \pi  \ell$.

The vacuum $|0 \rangle $ is a pure state
of this CFT. However a single observer can probe features of
this state
only within the observer horizons, i.e. in the diamond
region covered by the coordinates \rtzy. The results of
all such measurements are described by an observable density matrix
$\rho_{\rm obs}$. $\rho_{\rm obs}$ is constructed from the pure density matrix
$|0\rangle \langle 0|$ by tracing over the unobservable sector of the
Hilbert space supported behind the horizon. The entropy
\eqn\sent{S_{ent}=-tr\rho_{\rm obs} \ln \rho_{\rm obs} }
is nonzero because of correlations between the quantum states inside and
outside of the horizon. $S_{ent}$ is called the entanglement entropy
because it measures the extent to which the observable and unobservable
Hilbert spaces are
entangled. Note that the entanglement entropy, defined this way, agrees 
with the entropy of the gas of particles at the local Rindler temperature. 
 A general formula for $S_{ent}$ was derived in
\refs{\ch, \fpst}:
\eqn\snt{S_{ent}= {c \over 6}\rho|_{boundary} -{c\Delta \over 6},}
where $\Delta$ is the  short distance cutoff
and $\rho$ is the conformal factor of the metric in the coordinates (in our
case \dss\ ) used to define the vacuum evaluated at the boundary
(consisting of two points) of the
unobserved region. From \dss\ we see that the boundary is at
$t=0$, so $\rho = \ln \ell$. Putting this all together and using
$c=3M_pL$ we get
\eqn\sdnt{S_{ent}= LM_p \ln \ell,}
in agreement with  \sfa.

This result is a generalization to the two-dimensional deSitter case of
the observation of \fpst\ that, in a two-dimensional theory in which the
entirety of Newton's constant is induced from matter, the
Bekenstein-Hawking black hole entropy can be microscopically derived as
entanglement entropy.  The missing ingredient in
both of these previous discussions was a motivation for the assumption
that Newton's constant is induced. Here we see it is
natural - or at least equivalent to other assumptions -
in the brane-world context.

\newsec{Four Dimensions}

We can also consider a four dimensional brane world model. 
We have a four dimensional brane bounding two $AdS_5$ regions. 
If we consider perturbations of the four dimensional metric we
can analyze the system by first finding a five dimensional solution
which has a given four dimensional metric at the brane. The solution
will look like
\eqn\fdgrv{ ds^2 = L^2 \left( { g_{ \mu \nu}(z,x) dx^\mu
dx^\nu  + dz^2 \over z^2 } \right) + ... ~~~~~~ z\geq \epsilon }
where $g_{\mu\nu}(z=\epsilon,x) = g_{4~\mu\nu}$ is the four dimensional
metric. We can then insert the solution with a given four dimensional
metric back  into the action and get an effective action for the 
four dimensional metric. This effective action will contain a leading 
term going like $1/\epsilon^4$ which will be canceled by the brane
tension so that the next nontrivial term will be 
\eqn\fdgrav{
S(g_4) =  { 2 \over 16\pi G_5} \int_{z \geq \epsilon}  d^5x \sqrt{g_5} R_5 =
 { L^3 \over 16  \pi G_5 \epsilon^2 } \int d^4x \sqrt{g_4} R_4,
}
where to this order of approximation we can assume that the five 
dimensional metric is independent of $z$ (the $z$ dependent parts give
terms going like lower powers of $1/\epsilon$) and we took  into 
account the two copies of $AdS_5$.    
This of course the way that the four dimensional Newton's constant is 
computed in \rs.\foot{ In \rs\ $\epsilon$ does not appear since 
the four dimensional metric is rescaled by $L^2 \over \epsilon^2$.} 
We have phrased the calculation in this way to make connection with
AdS/CFT \refs{\jm,\suwi,\pp}, so that the integral over five dimensional 
metrics with the boundary metric at $z=\epsilon$ held fixed 
can be interpreted 
as a field theory with a cutoff $\epsilon$ on that particular four 
dimensional space. So the physics of \rs\ is the same as the physics
of a conformal field theory with a cutoff coupled to four dimensional
gravity (as considered in \vf ), where the four dimensional conformal
field theory as an $AdS$ dual, (see also \refs{\gub,\witten}).
Notice that the four dimensional Newton constant can be written as 
\eqn\fdnew{
{1 \over G_4} = {8  N_{dof} \over \pi \epsilon^2 }~,~~~~~~ N_{dof} \equiv
{ \pi L^3
\over 8 G_5 }
}
where $N_{dof}$ is the quantity that appears in all AdS/CFT calculations
involving the stress tensor, calculations such as the two point
function of the stress tensor or the free energy at finite temperature,
etc. It can be viewed as the effective number of degrees of freedom
of the CFT, ($N_{dof} = N^2/4$ for ${\cal N} = 4 $ SYM).
 This form for the four dimensional Newton constant is very
suggestive. It is of the general form expected for induced gravity in
four dimensions. If we start in four dimensions with a theory with 
infinite or very large Newton constant and we integrate out the matter
fields we expect to get a four dimensional value for the Newton
constant which is rougly as in \fdnew  
\refs{\bkls, \mark, \ct,\su, \cw, \kb,\myers, \lw}. 
The precise value that we would get seems to depend on the cutoff
procedure.
Indeed, if we use heat kernel regularization we would get that for
$N=4$ Yang Mills this cuadratic divergence cancels. 
The  gravity procedure of fixing the boundary at some finite 
distance must correspond to a suitable cutoff for the field theory and
it is not obvious that we should get the same results for divergent terms. 
Indeed, the supergravity regularization procedure would also
give a divergent value for the vacuum energy (which is being cancelled by
the brane). 
Again in theories where the four dimensional Newton constant is induced 
one can interpret black hole entropy as entanglement entropy \tj. 
If we consider a four dimensional metric with a horizon, like a black hole
or de-Sitter space we indeed find that the entropy is given by 
\eqn\entr{
S = {A_4 \over 4 G_4} = {2  N_{dof}  A_4  \over \epsilon^2 \pi  }
}
where we just used the relation of the 4d Newton constant and the
four dimensional parameters. The right hand side can be interpreted
as entanglement entropy. In other words, we can compute the 
entanglement entropy in the field theory as entropy of the 
gas of particles in thermal Rindler space and we would obtain
precisely the right hand side of \entr . We can do the entropy 
 calculation
at weak coupling in weakly coupled ${\cal N} =4$ SYM  and we would obtain 
agreement up to a numerical factor, which could be be related to
the ignorance of the cutoff procedure, but more fundamentally can 
also be related to strong coupling effects like the $3/4$ appearing in the 
relation between the weakly coupled and the strongly coupled expressions
for the free energy.

\newsec{Black Hole Formation on the Brane}

The entropy of a bulk black hole in the interior of AdS can be  
accounted for by representing it as a thermal state in the brane 
theory on its boundary. At first this may seem to be at odds with the 
accounting given here of the entropy of a black hole on the 
brane in terms of quantum entanglement. In this section we will 
attempt to reconcile the accounts by throwing a bulk black hole at the
brane and watching it turn into a brane black hole.\foot{Similar
ideas are being pursued by H. Verlinde.}

Consider a bulk black hole at the origin of AdS at temperature 
$T_H$ whose size is large
compared to the AdS radius $L$ but small compared to the brane radius 
$\ell$. This has a stable ground state in which there is a cloud of 
thermal radiation surrounding the black hole. In the brane theory, 
this is represented as homogeneous thermal state at temperature $T_H$. 
The statistical entropy of this state agrees with Bekenstein-Hawking 
entropy of the black hole.\foot{ The equality is precise for 
AdS$_3$. In higher dimensional cases such as AdS$_5$ 
it follows if one accepts 
the factor of $4/3$ as a feature of strongly coupled gauge theory, 
which we shall for the purposes of this discussion.}

The center of mass of the black hole can be given momentum by 
the action of an AdS$_{D+1}$ $SO(D,2)$ isometry. These isometries are broken
by the 
presence of the brane, but if black hole is not too near the brane 
this should not matter. The $SO(D,2)$  action will 
impart momenta to the black hole and make it
oscillate about the origin. The brane version of such a state 
can be found by applying an $SO(D,2)$ conformal transformation to the thermal 
brane state. The resulting state will carry conformal charges and 
have energy densities with bipolar oscillations. 
The statistical entropy of this 
oscillating state will of course still agree with Bekenstein-Hawking 
entropy of the oscillating black hole.   

In the above discussion we implicitly assumed that the field theory was defined 
on the cylinder ($S^3 \times R$).
 When we think of the field theory defined in 
flat space or de-Sitter space we are looking only at some coordinate 
patch of the AdS cylinder.  
In this case we only see half a period of oscillation which can be
interpreted
as a gas of particles in the field theory that contracts and expands
again. 

If the oscillation is  made large enough, the black hole actually reaches
the brane where it will stick, at least temporarily. The brane picture of
this process is that the oscillations in the energy density have become so
large that the thermal radiation collapses to from a black 
hole.\foot{Note that the 
coupling to gravity breaks conformal invariance so the conformal charges
are not conserved when the collapse occurs.}

Before collapse, the entropy is accounted for on the brane as 
the entropy of thermal radiation. After collapse, it is 
accounted for by the thermal gas of Rindler particles near the 
horizon.  (In general the black hole formation 
is not adiabatic.) 
This latter entropy is localized within a distance of the order of the 
cutoff from the horizon.  This could be described by saying that 
all stages the entropy is stored in thermal radiation, and this 
radiation hovers 
outside the horizon when the black hole is formed. In this description the 
statistical origin of the entropy of bulk and brane black holes 
appears to be similar. 
Eventually the black hole will evaporate and the final state will be just
outgoing thermal radiation on the brane theory.

It is interesting that there is a ``correspondence principle'' in the 
sense that when the AdS black hole has a radius of the order of 
the five dimensional anti-de-Sitter space and it makes a grazing collision
with the brane, then the entropies calculated as a thermal gas  and  as
entanglement entropy  are the same up to a numerical constant.
Such a black hole would have a
Schwarschild
radius in the boundary theory equal to the field theory cutoff $\epsilon$.

In closing, it  
remains to find a fully consistent quantum realization of such
a brane-world scenario to which our observations can be applied.
Alternately, perhaps it is applicable in a more general
setting. One of the important lessons of string duality is that 
something which is classical from one point of view can be quantum from 
another. What is needed here is a point of view from which
Newton's constant - usually regarded as a largely classical quantity - 
is a purely quantum effect. It is notable in this regard that closed
string poles arise as a loop effect in open string theory, indicating
that there might be a way in which the full closed string dynamics is
contained in open string field theory. In that case 
Newton's constant could be induced.

\centerline{\bf Acknowledgements}

This work was supported in part by DOE grant DE-FG02-91ER40654.
J.M. was also supported in part by the Sloan and Packard fellowships. 
We would like  to thank T. Jacobson, A. Tseytlin and H. Verlinde 
 for discussions.

\listrefs

 \end